\documentstyle[12pt]{article}

\newcommand{\beq}{\begin{equation}}
\newcommand{\eeq}{\end{equation}}
\newcommand{\beqa}{\begin{eqnarray}}
\newcommand{\eeqa}{\end{eqnarray}}
\newcommand{\beqar}{\begin{eqnarray*}}
\newcommand{\eeqar}{\end{eqnarray*}}

\newcommand{\eps}{\epsilon}
\newcommand{\g}{\gamma}
\newcommand{\G}{\Gamma}
\newcommand{\ka}{\kappa}

\renewcommand{\l}{\lambda}

\newcommand{\sig}{\sigma}

\newcommand{\z}{\zeta}
\newcommand{\inn}{\!\cdot\!}

\newcommand{\eg}{{\it e.g.,}\ }
\newcommand{\ie}{{\it i.e.,}\ }
\newcommand{\labell}[1]{\label{#1}} 
\newcommand{\reef}[1]{(\ref{#1})}
\newcommand\prt{\partial}
\newcommand\T{\tau}
\newcommand\veps{\varepsilon}
\newcommand\ls{\ell_s}
\newcommand\cF{{\cal F}}
\newcommand\cL{{\cal L}}
\newcommand\cG{{\cal G}}

\newcommand\bz{\bar{z}}
\newcommand\bF{\bar{F}}
\newcommand\bc{\bar{c}}

\newcommand\tG{{\tilde G}}
\newcommand\tV{\tilde V}
\newcommand\tg{{\tilde g}}
\newcommand\tS{\tilde S}
\newcommand\tb{\tilde b}

\newcommand\tB{{\widetilde B}}

\newcommand\Tr{{\rm Tr}}

\begin{document}

\thispagestyle{empty}
\rightline{\small hep-th/9901085 \hfill IPM/P-99/5}
\vspace*{1cm}

\begin{center}
{\bf \Large String Scattering from D-branes\\ [.25em]
in Type 0 Theories }
\vspace*{1cm}

{Mohammad R. Garousi\footnote{E-mail:
mohammad@physics.ipm.ac.ir and biruni@dci.iran.com}}\\
\vspace*{0.2cm}
{\it Institute for Studies in Theoretical Physics and Mathematics IPM} \\
{P.O. Box 19395-5746, Tehran, Iran}\\
\vspace*{0.1cm}
and\\
\vspace*{0.1cm}
{\it Department of Physics, Birjand University, Birjand, Iran}\\
\vspace*{2cm}
ABSTRACT
\end{center}
We derive fully covariant expressions for all two-point 
scattering amplitudes involving closed string tachyon and massless  strings
from Dirichlet brane in type 0 theories.
The amplitude for two massless D-brane fluctuations to produce closed
string tachyon is also evaluated.
We then examine in detail these  string scattering amplitudes in order
to extract world-volume couplings of the tachyon with itself and  with massless fields 
on a D-brane. 
We find  that the tachyon  appears as an overall coupling function in the 
Born-Infeld action and conjecture form of the function. 
\vfill
\setcounter{page}{0}
\setcounter{footnote}{0}
\newpage

\section{Introduction} \label{intro}

Recent years have seem dramatic progress in the understanding
of non-perturbative aspects of string theory\cite{excite}.
With these studies has come the realization that extended
objects, other than just strings, play an essential role.
An important tool in these investigations has been Dirichlet brane\cite{tasi}.
D-branes are non-perturbative states on which open string can live, and to which
various closed string states can couple.

In perturbative type II superstring theories, scattering amplitudes describing
the interaction of two massless closed string states with unexcited D-branes,
and one massless closed string with two open strings were studied
extensively in \cite{scatc, ours} and \cite{aki,scatd}, respectively. Perturbative  spectrum of bosonic
type 0 theories have the same massless states as the bosonic part of type II,
but with the doubled set of Ramond-Ramond(RR) states\footnote{Sectors of type 0 theories
are
\beqa
{\rm type \,0A}:&&\qquad (NS_-,NS_-)\oplus(NS_+,NS_+)\oplus(R_+,R_-)\oplus(R_-,R_+)\nonumber\\
{\rm type \, 0B}:&&\qquad (NS_-,NS_-)\oplus(NS_+,NS_+)\oplus(R_+,R_+)\oplus(R_-,R_-)
\nonumber
\eeqa
}\cite{dixon}. Hence 
the elementary D-branes  of the theory carry simultaneously two
RR charges\cite{bergman,kell}.
One can use the same techniques as in type II theories to
evaluate various string scattering with the D-branes in type 0 theories. Most
of these amplitudes are already presented in above citations. At the massless
level, scattering amplitude describing interaction of D-brane with two different RR states
is the only amplitude that should be evaluated.
The bosonic type 0 theories have closed string tachyon($\T$) in their spectrum
as well\cite{dixon}. Therefore, scattering amplitudes involving this tachyon can not
be address from previous work on type II theories either. The present paper
provide calculations on these amplitudes.

The paper is organized as follows. In the following section we describe the 
calculations of the scattering of two closed string tachyon, one tachyon 
and one closed string massless state, and two different massless RR states from D-brane
using conformal field theory techniques. The open string poles of these amplitudes
are consistent with the GSO projection of the open strings attached to the the D-branes
\cite{polyakov,kell}. In section 3 we evaluate the scattering amplitude of one
closed string tachyon and two open string states. Then, in section 4 we examine in detail these 
string amplitudes in order to find the world-volume coupling of tachyon to itself and to the massless fields on the D-branes.
We find that the tachyon does not have world-volume coupling  to RR fields.
However, the couplings of tachyon to other massless fields and to itself are
consistent with an extension of the Born-Infeld action. In particular, the tachyon
appears as an overall coupling function in the Born-Infeld action. We provide more evidence in support of
this action by studying some simple string amplitudes using the conformal field
theory with background fields.
We conclude with a brief discussion of our results in section 5.

Before continuing with our calculations, let us make a comment on
conventions. In the scattering amplitudes below, we will
set $\ls^2=\alpha'=2$. Our index conventions are that lowercase Greek
indices label vector in the entire ten-dimensional
spacetime, \eg $\mu,\nu=0,1,\ldots,9$; early Latin indices label vector
in the world-volume, \eg $a,b,c=0,1,\ldots,p$; middle Latin indices
label vector in the transverse space, \eg $i,j =p+1,\ldots,8,9$; early capital
Latin indices label spinor in the ten-dimensional spacetime, 
$\eg A,B,C=1,2,\cdots 32$; and middle capital Latin indices label
vector in $U(N)$ gauge space, $\eg I,J=1,2,\cdots ,N^2$.

\section{Two-point amplitudes} \label{scat}

Scattering amplitudes describing interaction  of strings with a fixed D-brane 
at tree level are calculated as the string vertex operator insertions
on a  disk (world-sheet) with Dirichlet and Neumann boundary conditions. In conformal
field theory frame, effect of these boundary conditions appears in  the 
world-sheet propagators. Alternatively, one may use a doubling trick to
convert the propagator to the standard form of open string propagator 
and shift the effect to the momentum
and polarization of vertex operators\cite{ours}. Hence, in order to find scattering amplitude
of strings with D-brane,  one may replace these modified
momenta and polarizations into the appropriate known scattering amplitudes of open
strings\cite{ours,aki}. In \cite{ours,aki}, using this idea, the authors were able to find
various massless superstring scattering with D-brane in type II theory by
relating them to the known amplitudes of type I theory. However,  type I theory
does not have open string tachyon, so the 
scattering amplitudes describing interaction of the closed string
tachyon with D-branes in type 0 are not related  to amplitudes in type I theory.

In  all the string scattering amplitudes appearing in this paper, we use the doubling trick 
and then  explicitly evaluate correlation
functions and take the integrals appearing in the amplitudes.

\subsection{Tachyon and Tachyon  amplitude} \label{tt}

We begin by calculating the amplitude describing the scattering of two closed
string tachyon from a D-brane. It is given by the following correlation function:
\beqa
A^{\T,\T}&\sim&\int d^2z_1d^2z_2\,<V^{\T}(p_1,z_1,\bz_1)\,V^{\T}(p_2,z_2,\bz_2)>
\labell{att}
\eeqa
where the tachyon vertex operators are 
\beqa
V^{\T}(p_1,z_1,\bz_1)&=&:V_0(p_1,z_1)::V_0(p_1\inn D,\bz_1):
\labell{v0}\\
V^{\T}(p_2,z_2,\bz_2)&=&:V_{-1}(p_2,z_2)::V_{-1}(p_2\inn D,\bz_2):
\labell{v1}\nonumber
\eeqa
where $p_i^2=1$ and $D^{\mu}{}_{\nu}$ is a diagonal matrix with
values +1 on the world-volume and -1 in the transverse space \cite{ours}.
In the vertex operators, subscripts 0 and -1 refer to their superghost charges
 which
must be added to -2 for the disk world-sheet. Here we have also used the doubling
trick to write the closed string vertex operator in terms of only holomorphic components.
The holomorphic components $V_0$ and $V_{-1}$ are given by
\[
V_0(p,z)\,=\,ip\inn \psi(z)\,e^{ip\cdot X(z)}\,\,\,\,\,\,\,\,
V_{-1}(p,z)\,=\,e^{-\phi(z)}e^{ip\cdot X(z)}
\]
Using the standard world-sheet propagators of type I theory, one
can evaluate the correlations in \reef{att} and show that the integrand is 
SL(2,R) invariant. We refer the reader to ref.~\cite{ours} for the details of
fixing this symmetry and performing the integrals, and simply state the 
final result
\beqa
A(\T,\T)&=&-i\frac{\ka^2 T_p}{2}\frac{\Gamma(-t/2)\Gamma(-2s)}{\Gamma(-1-t/2-2s)}
\labell{Att}
\eeqa
where $t=-(p_1+p_2)^2=-2-2p_1\inn p_2$ is the momentum transfer to the D-brane
and $s=-(p_{1a}p_1^a)=-(p_{2a}p_2^a)$ is the momentum flowing parallel to the 
world-volume of the brane. We also normalized \reef{Att} and subsequent 
amplitudes in this section 
by $-i\ka^2 T_p/2$, the same factor
as for the massless  closed string
scattering amplitudes\footnote{Note that there is a change in conventions
between \cite{ours} and the present 
paper: $(T_p)_{\rm there}=(\ka T_p)_{\rm here}$.} \cite{ours}.

From the gamma function factors in \reef{Att},
we see that the amplitude contain two infinite series poles corresponding
to closed and open string states in $t$-channel and $s$-channel, respectively, with
\[
m_{closed}^2\,=\,4n/{\alpha'}\,\,\,\,\,,\,\,\,\,\,m_{open}^2\,=\,m/{\alpha'}
\]
where $n,m=0,1,2,\cdots$. The closed string states in the $t$-channel belong only to
the $(NS_+,NS_+)$ sector of the theory. Here one concludes that either the coupling of the $(NS_-,NS_-)$ states to the D-branes or to the two tachyons in the bulk  is zero. From the amplitudes in sections 2.2 and 2.4, one finds that the $(NS_-,NS_-)$ stat
es do couple with the D-branes. Hence one concludes that the coupling of two
tachyons and the $(NS_-,NS_-)$ states is zero, \eg coupling of three tachyons
is zero in the bulk space \cite{kell}.  
The open strings in the $s$-channel are in the $NS_+$ sector. This is consistent with the GSO projection for the open string states on the D-brane \cite{kell}. However this amplitude does not rule out the $NS_-$ sector,  because the coupling
of closed string tachyon and $NS_-$ states is zero, \eg because of the world-sheet fermions in the vertex operator of open string tachyon, the world-volume coupling of open and closed tachyon is zero.

\subsection{Tachyon and NS-NS amplitudes}\label{tnsns}

Next, we evaluate the amplitudes describing the scattering of one tachyon and
one massless NS-NS(graviton, dilaton or Kalb-Ramond(antisymmetric tensor))
 states from the D-branes. The amplitude is given by
\beqa
A^{\T,NSNS}&\sim&\int\,d^2z_1d^2z_2\,<V^{\T}(p_1,z_1,\bz_1)V^{NSNS}(\veps_2,p_2,z_2,\bz_2)>
\labell{atnsns}\nonumber
\eeqa
where the tachyon vertex operator is given in \reef{v0} and the NS-NS vertex 
operator is
\beqa
V^{NSNS}(\veps_2,p_2,z_2,\bz_2)=(\veps_2\inn D)_{\mu\nu}:V^{\mu}_{-1}
(p_2,z_2)::V^{\nu}_{-1}(p_2\inn D,\bz_2):
\labell{vnsns}\nonumber
\eeqa
where $p_2^2=0$. Here $\veps_2^{\mu\nu}$ is the polarization of NS-NS 
states which is traceless and symmetric
(antisymmetric) for graviton(Kalb-Ramond) and 
\beqa
\veps_2^{\mu\nu}&=&\frac{1}{\sqrt{8}}(\eta^{\mu\nu}-\ell^{\mu}p_2^{\nu}
-\ell^{\nu}p_2^{\mu})\,\,\,\,\,\,\,\,;\,\,\,\,\,\,\,\ell\inn p_2=1
\labell{veps}
\eeqa
for dilaton. Again, we refer the reader to refs.~\cite{scatc,ours,scatd} for the details of the calculations.
The final result in this case is
\beqa
A&=&-\frac{i\ka^2 T_p}{2}\left(\Tr(\veps_2\inn D)\frac{\G(-t/2+1/2)\G(-2s)}
{\G(-1/2-t/2-2s)}\right.\labell{Atnsns}\\
&&\left.+p_1\inn\veps_2\inn p_1\frac{\G(-1/2-t/2)\G(-2s)}
{\G(-1/2-t/2-2s)}
-p_1\inn D\inn\veps_2\inn D\inn p_1
\frac{\G(-t/2+1/2)\G(-2s)}{\G(1/2-t/2-2s)}\right)
\nonumber
\eeqa
where $t=-(p_1+p_2)^2=-1-2p_1\inn p_2$. 
As a check of our calculations, we  inserted the dilaton polarization \reef{veps}
into \reef{Atnsns} and found that, as expected,   
the auxiliary vector $\ell^{\mu}$ disappears from the amplitude. 

The $t$- and 
$s$-channel poles of the  amplitude indicates that 
the mass of the closed and open string
in these channels are
\[
m_{closed}^2\,=\,(4n-2)/{\alpha'}\,\,\,\,\,\,,\,\,\,\,\,\,m_{open}^2\,=\,m/{\alpha'}\,\,.
\]
The closed strings in $t$-channel belong to the $(NS_-,NS_-)$ sector. This indicates that closed string state in this sector couple to the D-branes, \eg the closed string tachyon couple to the branes \cite{kell}. The poles of $t$-channel also indicate tha
t the coupling of one tachyon, one massless NS-NS state and any of the $(NS_+,NS_+)$ states is zero in the bulk, \eg the coupling of two massless NS-NS and one tachyon is zero in the bulk space \cite{kell}. Again the open strings in the $s$-channel is con
sistent with the GSO projection.

\subsection{Tachyon and RR amplitude}\label{trr}

Type 0 theories have two set of RR states, each made of direct
product of two open string state in  R sector. At the massless level, they are
\beqa
{\rm type\, 0A}:&&\quad{\bf (8_-\otimes 8_+)\oplus(8_+\otimes 8_-)\,=\,
(8_1+56_3)+(8_1+56_3)}\nonumber\\
{\rm type\, 0B}:&&\quad{\bf (8_-\otimes 8_-)\oplus(8_+\otimes 8_+)\,=\,
(1_0+28_2+35_4^{(+)})+(1_0+28_2+35_4^{(-)})}\nonumber
\eeqa
where symbols  indicate the irreducible 
representations and
the subscripts 0,1,2,3 and 4 indicate a scalar, vector, antisymmetric 2-, 3-
and 4-tensor, respectively. Here, the subscript $(+)((-))$  means that for
this tensor the field strength satisfies a self(anti-self)-duality condition.
The vertex operator corresponding to each of these states is also tensor
product of two open string vertex operators  with an appropriate polarization tensor.
Scattering amplitude of one of these states, the tachyon and  D-brane is given
by
\beqa
A^{\T,RR}&\sim&\int\, d^2z_1d^2z_2\,<V^{\T}(p_1,z_1,\bz_1)V^{RR}(\veps_2,p_2,z_2,\bz_2)>
\labell{atrr}
\eeqa
where the RR and tachyon vertex operators are 
\beqa
V^{RR}(\veps_2,p_2,z_2,\bz_2)&=&(P_{\mp}\G_{2(n)}M_p)^{AB}:V_{-1/2A}(p_2,z_2):
:V_{-1/2B}(p_2\inn D,\bz_2):\labell{vrr}\\
V^{\T}(p_1,z_1)&=&:V_0(p_1,z_1)::V_{-1}(p_1\inn D, \bz_1):\,\,\,.
\nonumber
\eeqa
The two different chiral projection operators, $P_{\mp}$, refers
to the two different set of RR states ($c$ and $\bc$). 
Here we have chosen the holomorphic and 
anti-holomorphic parts of the tachyon vertex operator in two different
picture in order for saturating the superghost charge of the world-sheet.
We refer the reader to ref.~\cite{ours}
for our conventions. In evaluating the correlators in \reef{atrr}, one
needs the correlation of two right-handed or left-handed spin operators
and one world-sheet
fermion that is given by(see, \eg  \cite{polchinski})
\beqa
<:S_A(z_1)::S_B(z_2)::\psi^{\mu}:>&=&2^{-1/2}(\g^{\mu})_{AB}z_{12}^{-3/4}
(z_{13}z_{23})^{-1/2}\,\,.\nonumber
\eeqa
Other correlators in \reef{atrr} can easily be evaluated. The final result is
\beqa
A(\T,c)&=&\frac{\ka^2 T_p}{2\sqrt{2}}\Tr(P_-\G_{2(n)}M_p\g\inn p_1)
\frac{\G(-t/2)\G(-2s)}{\G(-t/2-2s)}
\labell{Atrr}\nonumber
\eeqa
and an identical amplitude for $A(\T,\bc)$. Poles of this amplitude are at 
\[
m_{closed}^2\,=\,4n/{\alpha'}\,\,\,\,\,\,,\,\,\,\,\,\,
m_{open}^2\,=\,m/{\alpha'}
\]
The $t$-channel poles indicate that there is no coupling between one tachyon,
one massless RR and one of the $(NS_-,NS_-)$ states, \eg 
there is 
no coupling between two tachyons and one massless $RR$ field in the bulk space. 
Moreover, the massless pole
in this channel is consistent with having the coupling  $\T c\bc$ in the bulk\cite{kell}.
The massless pole in $s$-channel means that the tachyon couple to the open string
scalar fields
on the world-volume of the D-brane.

\subsection{$RR$ and $RR$ amplitude}\label{rrrr}

The amplitudes describing the scattering of two $c$'s or two $\bc$'s with D-brane
is exactly the same as the corresponding amplitudes in type II theories\cite{ours}.
Whereas, the scattering amplitudes for $c$, $\bc$ and D-brane is different and
given by
\beqa
A^{c,\bc}&\sim&\int\,d^2z_1d^2z_2\,<V^{c}(\veps_1,p_1,z_1,\bz_1)
V^{\bc}(\veps_2
,p_2,z_2,\bz_2)>
\labell{RRRR}
\eeqa
where the vertex operators are given in \reef{vrr}. 
In this case, in order to calculate the
amplitude \reef{RRRR}, 
one needs the correlation function between two right-handed and two
left-handed spin operators which is given by (see, \eg \cite{polchinski})
\beqa
<S_A(z_1)S_B(z_2)S_{\dot{C}}(z_3)S_{\dot{D}}(z_4)>&=&
\frac{1}{2}(\g_{\mu})_{AB}(\g^{\mu})_{\dot{C}\dot{D}}
(z_{13}z_{14}z_{23}z_{24})^{-1/4}(z_{12}z_{34})^{-3/4}\nonumber\\
&&+C_{A\dot{C}}C_{B\dot{D}}(z_{12}z_{34})^{1/4}(z_{14}z_{23})^{-1/4}
(z_{13}z_{24})^{-5/4}\nonumber\\
&&-C_{A\dot{D}}C_{B\dot{C}}(z_{12}z_{34})^{1/4}
(z_{13}z_{24})^{-1/4}(z_{14}z_{23})^{-5/4}
\nonumber
\eeqa
where C is the charge conjugation matrix (see, \eg \cite{ours}). Using this correlator and performing the others, one arrives at the following
final result: 
\beqa
A(c,\bc)&=&-\frac{i\ka^2 T_p}{2}\left(\frac{1}{2}\Tr(P_-\G_{1(n)}M_p\g^{\mu})\Tr(P_+\G_{2(m)}M_p\g_{\mu})
\frac{\G(-t/2+1/2)\G(-2s)}{\G(1/2-t/2-2s)}\right.\nonumber\\
&&-\Tr(P_-\G_{1(n)}C^{-1}\G_{2(m)}^TC)\frac{\G(-t/2-1/2)\G(-2s+1)}{\G(1/2-t/2-2s)}
\nonumber\\
&&\left.-\Tr(p_-\G_{1(n)}M_p\G_{2(m)}M_p)\frac{\G(-t/2+1/2)\G(-2s+1)}{\G(3/2-t/2-2s)}\right)
\labell{Arrrr}
\eeqa
where $t=-(p_1+p_2)^2=-2p_1\inn p_2$. This amplitude has the pole structure 
\[
m_{closed}^2\,=\,(4n-2)/{\alpha'}\,\,\,\,\,\,,\,\,\,\,\,\,
m_{open}^2\,=\,m/{\alpha'}
\]
The closed string channel indicates that there is a coupling between $c$, $\bc$ and  $(NS_-,NS_-)$ states, \eg the $\T c\bc$
coupling in the bulk is non-zero. It also tell us that  the  coupling of $c$, $\bc$ and $(NS_+,NS_+)$ is zero, \eg there is no coupling between $c$, $\bc$ and
the massless NS-NS states in the bulk.

The amplitude \reef{Arrrr} is consistent with the GSO projection of the open strings, \ie
it has no pole corresponding to on-shell propagation of $NS_-$ strings. Alternatively,
the amplitude \reef{Arrrr} indicates that
there is no coupling between massless RR  and $NS_-$  strings on the D-brane
world-volume. For example, consider the coupling of the open string tachyon (a
state belonging to the $NS_-$ sector) to the RR state on the world-volume of the
D-brane. The coupling 
is given by
\beqa
A^{t,RR}&\sim&\int\,dxd^2z\,<V^t(2k,x)V^{RR}(\veps,p,z,\bz)>
\nonumber
\eeqa
where the open string tachyon vertex operator $V^t(2k,x)=V_{-1}(2k,x)$ 
and the RR vertex operator is given in \reef{vrr}. Performing the correlators
in above amplitude, one finds
\beqa
A(t,c)&\sim&F_{a_0\cdots a_p}(\eps^v)^{a_0\cdots a_p}
\nonumber
\eeqa
and similar result for $A(t,\bc)$ which are zero, because there is no $p$-form
RR potential in the type 0 theory. 

\section{Closed-Open-Open amplitudes}\label{coo}

Scattering amplitudes with two massless open strings and one closed string can
be read from the corresponding amplitudes in type II theories\cite{aki}. The amplitudes
describing interaction of one closed string tachyon and two massless open
strings is given by
\beqa
A^{NS,NS,\T}&\sim&\int\,dx_1dx_2d^2z\,\Tr<V^{NS}(\z_1,k_1,x_1)V^{NS}(\z_2,k_2,x_2)
V^{\T}(p_3,z_3)>
\nonumber
\eeqa
where the open string vertex operators are
\beqa
V^{NS}(\z_i,k_i,x_i)&=&\z_{i\mu}:V^{\mu}_{-1}(2k_i,x_i):T_i
\nonumber
\eeqa
for $i=1,2$ and the tachyon vertex operator is given by \reef{v0}. Here we
have defined $\z_{iI}T^I\equiv\z_{i}T_i$
where $T$'s are the $N\times N$ generators of the $U(N)$ gauge symmetry of the coincident D-branes.
Using appropriate world-sheet propagators from \cite{ours}, 
one can evaluate the 
correlations above and show that the integrand in $SL(2,R)$ invariant.
Gauging this symmetry by fixing  $z=i$ and $x_2=\infty$, one arrives at
\[
A\sim2^{-2s-2}\int dx_1\left( (2s+1)\z_1\inn\z_2-\frac{i\z_1\inn p_3\,\z_2\inn D
\inn p_3}{x_1-i}-\frac{i\z_1\inn D\inn p_3\,\z_2\inn p_3}{x_1+i}\right)
(x_1^2+1)^s\Tr[T_1T_2]\nonumber
\]
where the integral is taken from $-\infty$ to $+\infty$, and
$s=-p_{3a}p_3^a=-(k_1+k_2)^2=-2k_1\inn k_2$.
This integral is doable and the result is
\beqa
A&=&-\frac{i\ka}{2}\left(\frac{s}{2}\z_1\inn\z_2+p_3\inn\z_1\,p_3\inn D\inn \z_2+
1\leftrightarrow 2\right)\frac{\G(-2s)}{\G(-s+1)\G(-s)}\Tr[T_1T_2]
\labell{Ansnst}
\eeqa
We have also normalized the amplitude at this point by $-i\ka/{2\pi}$. Here
the D-brane coupling constant $T_p$ cancels with the normalization of open
string states \cite{garousi}.  This amplitude has the same pole structure 
as the corresponding amplitude for massless
closed string states\cite{aki}, \ie  $m_{open}^2=(2n+1)/{\alpha'}$.

\section{World-Volume interactions}\label{world}

To extract appropriate world-volume coupling of open and closed string states
from string theory amplitudes, one should 
expand the massive poles of  string amplitudes
and then write a field theory for massless fields and tachyon, if any,
to reproduce those results. In general, 
the field theory should contain infinite terms to 
reproduce effect of the massive poles of string amplitudes. However, at low energy,
one should not consider the
terms which have large number of momentum in the expansion. Using this idea, 
in \cite{garousi} we extracted world-volume interaction of different
massless open and closed string states from string amplitudes. We then showed that
they are consistent with Born-Infeld and Chern-Simons action provided the closed
string states treated as functionals of the nonabelian  scalar fields describing
transverse fluctuations of the D$p$-brane.
In the present case, the world-volume interaction of massless fields remain 
unchanged. The only difference is that now we have two set of RR fields.

Now we do the same calculation to extract world-volume interaction of tachyon
and massless fields. We start with the simple case of two open and one closed string
scattering amplitude. This amplitude \reef{Ansnst} has no massless pole, so at
low energy it gives only contact terms, that is,
\beqa
A(a,a,\T)&=&-\frac{i\ka}{4}\left(f_{1ab}f_2^{ab}\Tr[T_1T_2]+O(k^4)\right)
\labell{aaT}\\
A(\l,\l, \T)&=&\frac{i\ka}{2}\left((k_{1a}\z_{1i}k_2^a\z_2^i+
p_{3i}\z_1^ip_{3j}\z_2^j)\Tr[T_1T_2]+O(k^4)\right)
\labell{llT}
\eeqa
and $A(a,\l,\T)$ is zero. Here we have suggestively 
introduced $f_{iab}=i(k_{ia}\z_{ib}-k_{ib}\z_{ia})$.
These terms are reproduced in field theory by
\beqa
\cL&=&-\frac{\ka}{2}\Tr\left(T_p\T(X)+\frac{1}{2}(\prt\l)^2\T(X)+\frac{1}{4}(f)^2\/
\T(X)+\cdots\right)\, .
\labell{looc}
\eeqa
where $X^i=\l^i/{\sqrt{T_p}}$. Taylor expansion of the first term above  reproduces the second term of \reef{llT}, 
the second and last term of \reef{looc} reproduces the first term of equations
\reef{llT} and \reef{aaT}, respectively. The first term in \reef{looc} is consistent
with the result basing on the cylinder amplitude \cite{kell}.

Now to find non-derivative world-volume interaction of the tachyon 
to itself and to other massless closed string
fields, we expand their corresponding string amplitudes and ignore the  terms which have
two  and more momenta, that is,
\beqa
A(\T,\T)&=&-i\ka^2 T_p\left(\frac{1+2s}{t}-\frac{p_{1i}p_2^i}{4s}+\frac{3}{4}+
O(p^2)\right)
\labell{TT}\\
A(\T,h)&=&\frac{i\ka^2 T_p}{2}\left(\frac{2p_1\inn\veps_2\inn p_1}{t+1}+
\frac{p_{1i}p_2^i\veps_2^a{}_a+2p_{1i}\veps_2^{ia}p_{1a}}{s}-\veps_2^a{}_a+
O(p^2)\right)
\labell{Th}\nonumber\\
A(\T,\phi)&=&\frac{i\ka^2 T_p}{4\sqrt{2}}\left(\frac{2}{t+1}+
\frac{p_{1i}p_2^i(p-3)}{s}-(p-3)+O(p^2)\right)
\labell{Tphi}\nonumber
\eeqa
and $A(\T,b)$ is zero. Similar expansion can be done for $A(\T,c)$ and 
$A(c,\bc)$. These tachyonic and massless poles should be reproduced
in field theory by suitable world-volume and bulk actions in Einstein frame metric.
The appropriate actions in the world-volume are the Born-Infeld, the Chern-Simons
and eq.~\reef{looc}, and in the bulk 
is\footnote{Note that the fields in this paper have
the normalization of the vertex operators\cite{garousi}.}\cite{kell}
\beqa
S&=&\int d^{10}x\sqrt{-g}\left[\frac{1}{2\ka^2}R-
\frac{1}{2}(\nabla\phi)^2
-\frac{1}{2}(\nabla \T)^2+\frac{1}{2}\T^2 e^{\frac{\ka}{\sqrt{2}}\phi}
-\frac{16}{5!}F_5^{(+)}F_5^{(-)}f(2\ka\T)
\right.\nonumber\\
&&\left.\qquad\qquad-\frac{8}{n!}(F_{(n)}\inn F_{(n)}+\bF_{(n)}\inn\bF_{(n)}+4\ka F_{(n)}
\inn\bF_{(n)}\T)e^{\frac{\ka}{\sqrt{2}}(5-n)\phi}\right]
\labell{I}
\eeqa
where the tachyon coupling function is
\beqa
f(2\ka\T)&=&1+2\ka\T+\frac{1}{2}(2\ka\T)^2+\cdots\,\,.
\labell{fn}
\eeqa 
We use these actions to reproduce
the tachyon and massless
poles of the string amplitudes \reef{TT}, and then compare them with the
string theory results  to find the new contact terms
of the world-volume field theory. We will present the details of the calculations for scattering 
with two tachyons. We begin by calculating the  massless $t$-channel amplitude 
\beqa
A_t'(\T,\T)&=&i\tS_{\phi}\tG_{\phi}\tV_{\phi \T_1\T_2}+
i(\tS_h)^{\mu\nu}(\tG_h)_{\mu\nu}{}^{\alpha\beta}(\tV_{h\T_1\T_2})_{\alpha\beta}
\labell{tchan}
\eeqa
where we used the fact that in the bulk 
both dilaton and graviton couple to two tachyons.
Their corresponding  vertex functions
are derived from \reef{I}
\beqa
(\tV_{h\T_1\T_2})^{\mu\nu}&=&-2i\ka p_1^{(\mu}p_2^{\nu)}+i\ka\eta^{\mu\nu}(p_1\inn p_2+1)\nonumber\\
\tV_{\phi \T_1\T_2}&=&\frac{i\ka}{\sqrt{2}}\,\,.
\labell{vhtt}
\eeqa
the dilaton and graviton sources and their corresponding propagators can also be
derived from \reef{I}(they appear in \cite{ours}). Now substituting these into \reef{tchan}, one finds 
\beqa
A'_t(\T,\T)&=&-i\ka^2 T_p\left(\frac{1+2s}{t}\right)
\labell{A'tt}
\eeqa
The first term above reproduces the $t$-channel massless pole of string
amplitude \reef{TT}. This insures that the normalization of string amplitude \reef{Att}
is consistent with normalization of bulk action \reef{I}.
Now, the $s$-channel amplitude corresponding to propagation of massless 
open string on the 
D-brane world-volume can be evaluated as:
\beqa
A'_s(\T,\T)&=&\tV^i_{\T_1\l}\tG^{ij}_{\l}\tV^j_{\l \T_2}
\labell{Ass}\nonumber
\eeqa
where $\tG^{ij}_{\l}=iN^{ij}/s$ is the scalar propagator, and 
the vertex function is derived from the Taylor expiation of the first term in eq.~\reef{looc}
\beqa
\tV^j_{\l \T_k}&=&\frac{\ka}{2}\sqrt{T_p}p^j_k\,\, .
\nonumber
\eeqa
for $k=1,2$ and  $p^j_k$ is momentum of external tachyon. The $s$-channel amplitude then becomes
\beqa
A'_s(\T,\T)&=&i\ka^2 T_p\frac{p_{1i}p_2^i}{4s}\,\, .
\labell{A'ss}
\eeqa
This amplitude  reproduces the $s$-channel  pole of \reef{TT}. Hence, normalization
of \reef{looc} is consistent with \reef{Att} and subsequently with \reef{I}.
Now subtracting these field theory amplitudes 
\reef{A'tt} and \reef{A'ss} from \reef{TT}, one finds
\beqa
A(\T,\T)-A'_t(\T,\T)-A'_s(\T,\T)&=&-i\ka^2 T_p\left(\frac{3}{4}+O(p^2)\right)
\nonumber\\
A(\T,h)-A'_{t+1}(\T,h)-A'_s(\T,h)&=&-\frac{i\ka^2 T_p}{2}\left(\veps_{2a}{}^a+O(p^2)\right)
\nonumber\\
A(\T,\phi)-A'_{t+1}(\T,\phi)-A'_s(\T,\phi)&=&-i\ka^2 T_p\left(\frac{p-3}{4\sqrt{2}}+O(p^2)\right)
\nonumber\\
A(\T,c)-A'_t(\T,c)-A'_s(\T,c)&=&\ka^2 T_pO(p^2)
\nonumber\\
A(c,\bc)-A'_{t+1}(c,\bc)-A'_s(c,\bc)&=&\ka^2 T_pO(p^2)
\labell{contact}\nonumber
\eeqa
Here we have listed the results for the other closed string modes as well. 
These contact terms are reproduce in field theory by the following action
\beqa
\cL&=&-\frac{\ka^2 T_p}{2}\left(\T h^a{}_a+\frac{p-3}{2\sqrt{2}}
\T\phi+\frac{3}{4}\T^2+\cdots\right)
\labell{lcc}
\eeqa
Now, it is not difficult to see that eqs.~\reef{looc} and \reef{lcc} are consistent  
with the following generalized Born-Infeld action: 
\beq
S_{BI}^{\T}=-T_p \int d^{p+1}\sig\ \Tr\left(g(\ka\T)
e^{\frac{p-3}{2\sqrt{2}}\ka\phi}
\sqrt{-det(\tg_{ab}
-2\ka\tb_{ab}\,e^{-\frac{\ka}{\sqrt{2}}\phi}+
\frac{1}{\sqrt{T_p}}f_{ab}\,e^{-\frac{\ka}{\sqrt{2}}\phi})}\right)
\labell{biaction}
\eeq
where the closed string tachyon coupling function is
\beqa
g(\ka\T)&=&1+\frac{1}{2}\ka\T+\frac{3}{8}(\ka\T)^2+\cdots
\labell{ftach}
\eeqa
and the dots represents possible higher order terms which can not be read from
the scattering amplitudes considered in this paper. We refer the reader to \cite{garousi} for 
details of expanding the square root in eq.~\reef{biaction}.

Now we give more evidence in support of  eq.~\reef{biaction}. We evaluate some more
world-volume interactions in the presence of constant background field and show
that they are consistent with the generalized Born-Infeld action \reef{biaction}.
We consider constant background Kalb-Ramond field and (or) constant gauge field
strength, \ie, $\cF^{12}=-\cF^{21}\equiv\cF$. 
We begin with evaluating
the world-volume coupling of one closed string tachyon and one massless open string NS
state. 
At the world-sheet level, this coupling is given by
\beqa
A^{\T,NS}_{\cF}&\sim&\int dxdz\,\Tr<V^{\T}(p_1,\cF,z_1,\bz_1)V^{NS}(k_2,\z_2,\cF,x_2)>
\nonumber
\eeqa
Using the doubling trick, one may write
the tachyon and the open string vertex operators as
\beqa
V^{\T}(p_1,\cF,z_1,\bz_1)&=&:V_0(p_1,z_1)::V_{-1}(p_1\inn D_{\cF},\bz_1):\nonumber\\
V^{NS}(k_2,\z_2,\cF,x_2)&=&\z_2\inn \cG_{\mu}:V^{\mu}_{-1}(k_2+k_2\inn D_{\cF},x_2):T_2
\nonumber
\eeqa
where $\cG^{ab}=(\eta^{ab}+D_{\cF}^{ab})/2$ for the gauge fields and $\cG^{ij}=N^{ij}$
for scalar fields. And the $D_{\cF}^{ab}$ is a diagonal matrix 
given in \cite{mrg}. Performing
the correlators above, one finds
\beqa
A_{\cF}^{\T,NS}&\sim&\z_2\inn\cG\inn p_1\Tr[T_2]\,\,.
\nonumber
\eeqa
For the scalar and gauge states, it becomes
\beqa
A_{\cF}(\T,\l)&\sim&\z_{2i} p^i_1\Tr[T_2]\nonumber\\
A_{\cF}(\T
,a)&\sim&\frac{f_{2ab}\cF^{ab}}{1+\cF^2}\Tr[T_2]\nonumber
\eeqa
First term above is reproduced by Taylor expansion of the linear tachyon in \reef{biaction},
and the second term
by expanding the square root in  \reef{biaction} 
around the background field\cite{mrg}. This gives evidence that the tachyon
has linear coupling to the Born-Infeld terms. To show that the tachyon has quadratic
coupling to the Born-Infeld terms as well, we consider scattering amplitude 
of two tachyon and one massless open string NS state with D-brane. This amplitude
is given by 
\beqa
A_{\cF}^{\T,\T,NS}&\sim&\int dxd^2z_1d^2z_2\,\Tr<V^{\T}\,V^{\T}\,V^{NS}> \,\,.
\nonumber
\eeqa
We are not going to make efforts here to evaluate this amplitude explicitly. However, from
the Taylor expansion of the quadratic tachyon in \reef{lcc}, one may conclude
that the scattering amplitude above has the following contact term:
\beqa
A_{\cF}^{\T,\T,NS}&\sim&(\z_3\inn \cG\inn p_1+\z_3\inn \cG\inn p_2)\Tr[T_3]+\cdots
\nonumber
\eeqa
For the gauge fields, this amplitude becomes
\beqa
A_{\cF}(\T,\T,a)&\sim&\frac{f_{3ab}\cF^{ab}}{1+\cF^2}\Tr[T_3]+\cdots\,\,.
\nonumber
\eeqa
Again this contact term is reproduced by the action \reef{biaction}. 
Hence two tachyons couple with the entire Born-Infeld terms. This ends our evidences
in  support of the action \reef{biaction}.

\section{Discussion} \label{discuss}

In this paper we have evaluated the amplitudes describing interaction of two
closed string tachyons, one tachyon and one massless state, and two different
RR states with D-branes in type 0 theories. We have also evaluated scattering
amplitude of one closed string tachyon and two massless open string states.
We then examine these amplitudes in detail in order to extract their
world-volume couplings. We found that the tachyon dose not couple to the RR fields
on the D-branes. We have also found that the  coupling of tachyon to other massless fields
can be described by some extension of the Born-Infeld action \ie eq.~\reef{biaction}.

Evaluation of bulk effective actions for tachyon based on on-shell scattering amplitudes
calculations is uncertain. This steam from the fact that it is hard to distinguish
between $\T^2$ and $\T\prt_{\mu}\prt^{\mu}\T$, \ie in the scattering amplitude it becomes $p^2=1$. However, this is not the case for
the world-volume effective action. In this case, derivatives in the D-brane world-volume
 and the transverse spaces, \ie $\prt_a$ and $\prt_i$, appear differing in the
world-volume action. Hence the effective action \reef{biaction} is unambiguous.

From the tachyon coupling to the gauge and to the scalar fields in \reef{looc}, 
one concludes that this field  couples with the 10-dimensional Yang-Mills theory
reduced to the $p+1$ dimensional world-volume. Hence the action \reef{looc} or
\reef{biaction} is consistent with the T-duality.

The extended Born-Infeld action \reef{biaction} in terms of the standard 
normalized fields (\ie $2\ka\T=T$ and see \cite{garousi} for other fields) and in the string frame is
\beqa
S&=&-T_p\int d^{p+1}\sigma\,g(T/2)e^{-\Phi}\sqrt{-det(\tG_{ab}+\tB_{ab}+2\pi\ell_sF_{ab})}\,\,.
\nonumber
\eeqa
From this action, one finds that the effective Yang-Mills coupling 
is\footnote{In the original version of ref.~\cite{klts} the form of the Yang-Mills
coupling was conjectured to be $1/{g_{\rm YM}^2}\sim (1+T)e^{-\Phi}$.
However, later on they found independently that the coupling should have
structure $1/{g_{\rm YM}^2}\sim (1+T/4)e^{-\Phi}$.}
\beqa
\frac{1}{g_{\rm YM}^2}&\sim&(1+\frac{T}{4}+\frac{3T^2}{32}+\cdots)e^{-\Phi}
\nonumber
\eeqa
This is consistent with the power expansion of the following function
\beqa
\frac{1}{g_{\rm YM}^2}&\sim&\frac{e^{-\Phi}}{\sqrt{1-T/2}}
\nonumber
\eeqa
It would be interesting to find cubic coupling of tachyon to the D-branes and
see if it is consistent with the above conjectured function.

\vspace{1cm}

{\bf Acknowledgements}

I am grateful to M.H. Sarmadi and A.A. Tseytlin for very useful discussions. This work was supported by University of Birjand and IPM.

\end{document}